\documentclass[aps,prl,twocolumn,groupedaddress]{revtex4}
\usepackage{graphicx,amssymb}
\usepackage{dcolumn}
\usepackage{bm}

\begin{document}

\title{Quantum Metallicity on the High-Field Side of the
Superconductor-Insulator Transition}

\author{T.\,I.~Baturina$^{1,2}$\/\email{tatbat@isp.nsc.ru},
        C.~Strunk$^2$       }
\affiliation{ $^1$ Institute of Semiconductor Physics, 630090,
Novosibirsk, Russia\\ $^2$Institut f\"{u}r experimentelle und
angewandte Physik, Universit\"{a}t Regensburg, D-93025 Regensburg,
Germany}
\author{M.\,R.~Baklanov, A.~Satta}
\affiliation{IMEC, Kapeldreef 75, B-3001 Leuven, Belgium}

\date{\today}

\begin{abstract}
We investigate ultrathin superconducting TiN films, which are very close 
to the localization threshold. Perpendicular magnetic field drives 
the films from the superconducting to an insulating state, 
with very high resistance. Further increase of the magnetic field leads 
to an exponential decay of the resistance towards a finite value. 
In the limit of low temperatures, the saturation value 
can be very accurately extrapolated to the universal quantum resistance $h/e^2$. 
Our analysis suggests that at high magnetic fields a new ground state, 
distinct from the normal metallic state occurring above the superconducting 
transition temperature, is formed. A comparison with other studies 
on different materials indicates that the quantum metallic phase following 
the magnetic-field-induced insulating phase is a generic property of systems 
close to the disorder-driven superconductor-insulator transition.

\end{abstract}


\maketitle

The investigation of disordered superconducting films is of
fundamental importance to understand the impact of electron-electron
interaction and disorder on the ground state of many-body
systems~\cite{Goldman_Review}. Metal and insulator are two basic
ground states of the electrons in solids. The Cooper pairing, a
dramatic manifestation of the attractive part of the
electron-electron interaction, results in an instability of the
Fermi sea and the formation of a new ground state. This
superconducting state is characterized by long-range phase coherence
and the possibility of nondissipative charge transport. On the other
hand, disorder acts in opposite direction, as it favors the
repulsive part of the electron-electron interaction and the
localization of the electron wave function. The competition between
localization and superconductivity can result in an insulating
ground state - the so-called Bose-insulator, which is formed by
localized Cooper pairs~\cite{Fisher89,Fisher90_1,Fisher90_2}.

At zero temperature the transition between these two phases, the
superconductor-insulator transition (SIT), is driven purely by
quantum fluctuations and is one of the prime examples of a quantum
phase transition~\cite{Sondhi,Dolgop}. Experimentally, the SIT can
be induced by decreasing the film thickness~\cite{Haviland,Liu} and
close to the critical thickness also by magnetic
field~\cite{Hebard}. These possibilities are commonly distinguished
as disorder-driven SIT and magnetic-field driven SIT. In the latter
case, the magnetic field is supposed to suppress first the
macroscopic phase coherence, while the Cooper pairing may survive
locally. At sufficiently low temperatures this results in a sharp
increase of the resistance up to several orders of
magnitude~\cite{Shahar} - the Bose-insulator. Upon further increase
of the magnetic field, the localized Cooper pairs are gradually
destroyed, leading to a strongly negative
magnetoresistance~\cite{Destr}. So far, direct experimental evidence
for the existence of localized Cooper pairs is still scarce.

In agreement with the scenario above, early
investigations~\cite{Destr} of amorphous InO$_x$ films revealed a
strongly non-monotonic magnetoresistance. The resistance value in
the high-field limit roughly approached the normal state resistance
$R_N$, as expected, if the normal metal phase reappears. However, a
closer analysis of the data on amorphous
InO$_x$~\cite{Destr,SITVFG,KapitulnikSIT} and on our polycrystalline
TiN  in the limit $T=0$ and large $B $ reveals that the resistance
of the films does not return to $R_N$~\cite{OurTiN04}. Hence, the
nature of the Bose-insulator and its behavior in a strong magnetic
field remain an open issue.

In this Letter, we show that the magnetoresistance of ultrathin TiN
films decays exponentially at high magnetic fields and then
saturates at a value considerably {\it higher} than the normal state
resistance. The saturation resistance can be extrapolated with high
accuracy towards $T=0$ and turns out to be $h/e^2$, independent of
the degree of disorder. The application of our analysis to existing
data on InO$_x$ films \cite{SITVFG} reveals the same behavior. This
demonstrates the universal character of the theoretically so far
unexplained quantum metallic phase in disordered superconducting
films exposed to high magnetic fields.

The TiN films with a thickness of $\lesssim 5$~nm were formed by
atomic layer chemical vapor deposition onto a Si/SiO$_2$ substrate.
Structural analysis shows that the films consist of a dense packing
of crystallites, with a rather narrow distribution of sizes around
$\sim 30$~nm. The samples for the transport measurements were
patterned into Hall bridges using conventional UV lithography and
subsequent plasma etching. The film resistance was measured in
perpendicular magnetic field using a standard four-probe lock-in
technique. From earlier investigations, we estimated that the
product of Fermi wave vector and elastic mean free path $k_F\ell<1$
below 2~K \cite{OurTiN04}.

\begin{figure}
\includegraphics[width=80mm]{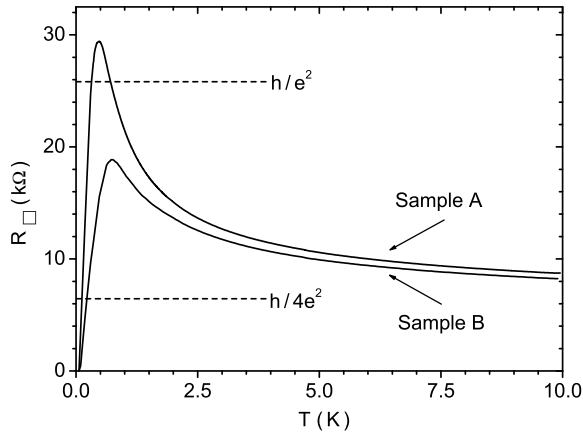}
\caption{\label{fig1:RTB0} Temperature dependence of $R_\square$:
The two TiN films differ slightly in their normal state resistance
$R_\square(10\,{\rm K})=8.24\,{\rm k}\Omega$ (sample A) and
8.74$\,$k$\Omega$ (sample B), respectively. The resistance reaches a
maximum at $R_\square (T_{\rm m}=0.48\,{\rm K})=29.4\,{\rm k}\Omega$
(sample A) and $R_\square (T_{\rm m}=0.72\,{\rm K})= 18.6\,{\rm
k}\Omega$ (sample B). All data in this work were taken at a
measurement frequency of $0.4 - 2$~Hz with an ac current $0.01 -
1$~nA. }
\end{figure}

In Fig.~\ref{fig1:RTB0}, we plot the temperature dependence of the
resistance per square $R_\square$ at zero magnetic field for two
film, which are very close to disorder-tuned SIT. As the temperature
decreases, $R_\square$ of both samples increases, then reaches a
maximum value at an intermediate temperature $T_{\rm m}$, and
finally drops again, showing the transition into the zero resistance
state. Note that there is no drop in the resistance around 5.6~K,
the bulk transition temperature of TiN. The latter would be a
characteristic of a granular film with only weak tunnel coupling
between the grains \cite{granular}. The absence of such a feature in
our data indicates that our films are nominally homogeneous with
strong metallic coupling between the crystallites
\cite{finkelstein}.

From the bosonic model, the critical resistance of the zero-field
superconductor-insulator transition is expected to be close to a
universal value - the quantum resistance for Cooper pairs
$h/(2e)^2$~\cite{Fisher89,Fisher90_1,Fisher90_2}. However, this is
still a controversial issue. Up to now only for Bi films
~\cite{Haviland,Liu}, a critical sheet resistance $R_c$ close to the
predicted value of $h/(2e)^2$ has been observed. In other materials
the resistance at the transition was found to deviate significantly
from the expected universal value, for instance, on Pb -- $R_c
\simeq 12$~k$\Omega$~\cite{Liu}, Al -- $R_c \simeq
24$~k$\Omega$~\cite{Liu}, Be -- $R_c \simeq
10$~k$\Omega$~\cite{Be3}. In sample A the maximal $R_{\square}$ at
$B=0$ even exceeds the value of $h/e^2=25.8$~k$\Omega$, implying
that the usual perturbative theories must fail to describe the data,
since the change of $R_\square$ is much larger than $R_\square$
itself.
\begin{figure}
\includegraphics{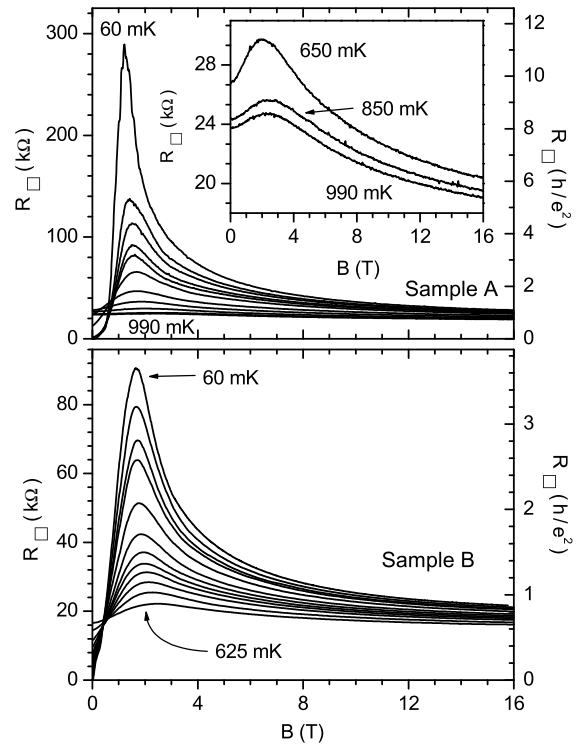}
\caption{\label{fig2:RBTiEXP} Sheet resistance $R_\square$ in
perpendicular magnetic field: Top: (sample A) $T=60$, 80, 95, 120,
140, 180, 300, 450, 650, 850, 990~mK. The inset shows a close-up
view of the $R(B)$-curves measured at temperatures corresponding
$dR/dT<0$ at zero magnetic field. Bottom: (sample B) $T=60$, 75, 90,
100, 130, 180, 220, 260, 300, 360, 480, 625~mK.
$R_\square(T=60\,{\rm mK})$ reaches a maximum at $B_\mathrm{m} \sim
1.2$~T (sample A) and at $B_\mathrm{m} \sim 1.6$~T (sample B). }
\end{figure}

In Fig.~\ref{fig2:RBTiEXP}, we show the magnetoresistance measured
at temperatures down to 60~mK and at magnetic fields up to 16~T. The
resistance varies nonmonotonically with $B$ and reaches a maximum
value at a magnetic field $B_\mathrm{m}$, followed by a rapid drop 
and gradual saturation at magnetic fields more than an order of
magnitude larger than $B_\mathrm{m}$. The value of $B_\mathrm{m}$
slightly shifts towards larger magnetic fields as the temperature
increases. Yet the nonmonotonic shape of $R(B)$ still persist even
when $T>T_\mathrm{m}$. This is seen in the inset of
Fig.~\ref{fig2:RBTiEXP} (top), where we plot magnetoresistance data
in the high temperature region corresponding $dR/dT<0$ at zero
magnetic field.

The saturation occurs at a resistance $R_\mathrm{sat}$ near the
quantum resistance $h/e^2$ (see right axes of
Fig.~\ref{fig2:RBTiEXP}). Interestingly, $R_\mathrm{sat}$ only
slightly increases as $T$ approaches zero. This indicates metallic,
rather than the insulating behavior observed at $B$ close to
$B_\mathrm{m}$.

\begin{figure}
\includegraphics[width=85mm]{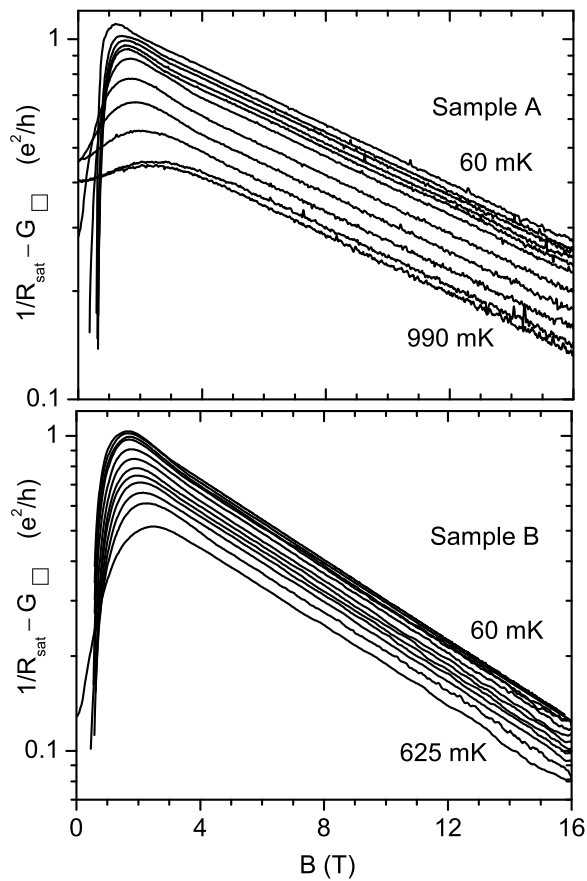}
\caption{\label{fig3:RBTiQM} Scaling plot of the data in
Fig.~\ref{fig2:RBTiEXP}: For certain values of $R_\mathrm{sat}$,
$\ln\left[1/R_\mathrm{sat}-G_\square(B)\right]$ varies linearly
vs.~$B$, with a $T$-independent slope. The linear slope corresponds
to a characteristic field $B^*\simeq 10.7$~T (sample A) and $\simeq
6.8$~T (sample B), respectively. }
\end{figure}
We now turn to the main result of this work, which is the analysis
of the negative magnetoresistance of our samples on the high-field
side of the superconductor-insulator transition, where the
saturation of $R(B)$ is observed (Fig.~\ref{fig2:RBTiEXP}). First,
we plotted the expression $\ln(1/R_\mathrm{sat} - G_\square(B))$
vs.~$B$, for each value of $T$. Here, $G_\square=1/R_\square$ is the
conductance per square. By varying the value of $R_\mathrm{sat}$ for
each curve, we could linearize $\ln(1/R_\mathrm{sat} -
G_\square(B))$ vs.~$B$ over a large range of $B$ with a
$T$-independent slope, as it is seen in Fig.~\ref{fig3:RBTiQM}. This
indicates that $G_\square\propto\exp(-B/B^*)$ exhibits a simple
exponential decay at high magnetic fields with a characteristic
magnetic field $B^*$. In addition, the curves in
Fig.~\ref{fig3:RBTiQM} show a slightly $T$-dependent offset.

\begin{figure}
\includegraphics[width=85mm]{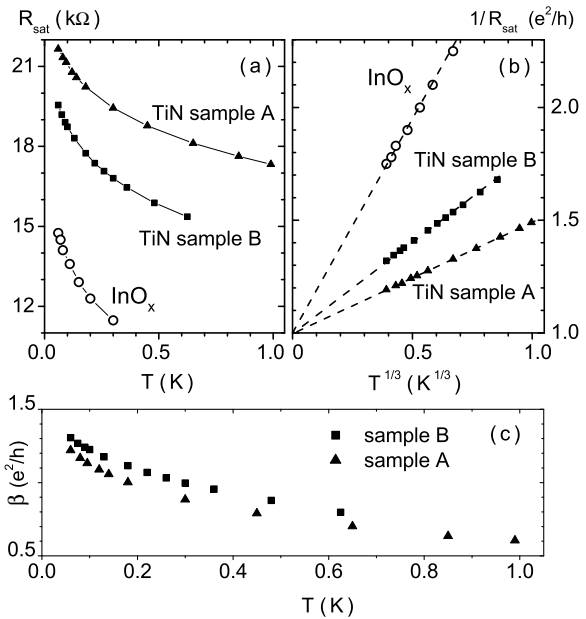}
\caption{\label{fig4:RsatBeta} Temperature dependence of
$R_\mathrm{sat}$ and $\beta$:~(a) $R_\mathrm{sat}(T)$, (b)
$R^{-1}_\mathrm{sat}(T^{1/3})$, and (c) $\beta(T)$ for the samples A
and B and the data from Fig.~1a of Ref.~\cite{SITVFG}. For the
latter, we obtained $B^*\simeq4.5\,$T and $\beta\simeq 2e^2/h$,
nearly independent of $T$.}
\end{figure}
These observations can be condensed into a simple phenomenological
expression for the high-field magnetoconductance as a function of
temperature and magnetic field:
\begin{equation}
G_\square(T,B) =1/R_\mathrm{sat}(T)- \beta(T)\exp(-B/B^*)\;,
\label{main}
\end{equation}
where $B^*$ is a constant, which increases with the degree of
disorder, and $\beta\simeq e^2/h$ is weakly $T$-dependent and
accounts for the slight offset of the curves for different $T$. The
$T$-dependence of $R_\mathrm{sat}$ and $\beta$ for both samples
under study is shown in Fig.~\ref{fig4:RsatBeta}(a) and (c),
respectively.

One more important result is the temperature dependence of
$R_\mathrm{sat}(T)$. For both samples, we have obtained a relatively
weak quasi-metallic $T$-dependence of $R_\mathrm{sat}$. A plot of
$1/R_\mathrm{sat}(T)$ vs.~$T^{1/3}$ (Fig.~\ref{fig4:RsatBeta}(b))
reveals that the $T$-dependent part of $1/R_\mathrm{sat}(T)$ closely
follows a $T^{1/3}$ power law. In addition, for both samples
$1/R_\mathrm{sat}(T)$ extrapolates in the limit $T\rightarrow 0$
very accurately towards $e^2/h$. This can already be seen in the
high-field region of $R(B)$ in Fig.~\ref{fig2:RBTiEXP} and indicates
a sample independent, possibly universal behavior. A nonmonotonic
magnetoresistance similar to our results was reported also for
disordered thin InO$_x$ films (see Fig.~1a in \cite{SITVFG} and
Fig.~3b in Ref.~\cite{KapitulnikSIT}). For the data of
Ref.~\cite{SITVFG} we have performed the same analysis as for our
data on TiN films and found the same scaling and the same
extrapolation $R_\mathrm{sat}(T\rightarrow 0)=h/e^2$. The specific
advantage of TiN over InO$_x$ is its significantly lower value of
$B_\mathrm{m}$ which allows the observation of
 $G_\square\propto\exp(-B/B^*)$ in a wide range of magnetic field~\cite{B_m}.

From these observations the following scenario emerges: the
superconducting state of our TiN films at $B=0$ is rapidly destroyed
by quantum phase fluctuations as the magnetic field is moderately
increased, while the Cooper pairing may survive locally. It was
recently suggested that strong mesoscopic fluctuations of the energy
gap \cite{feigelman05} can induce such phase fluctuations also in
{\it homogeneous} thin films. If localized Cooper-pairs and a
Bose-insulator exist, these are expected to be suppressed at higher
magnetic fields. This corresponds to a strong decrease of $R(B)$.
Phenomenologically, the exponential dependence of $G_\square(T,B)$
in Eq.~(1) may result from a broad dispersion of binding energies of
localized Cooper pairs~\cite{Larkin_pc}. As shown by Suzuki and
coworkers \cite{suzuki} there is little spin-orbit scattering in
TiN. Thus spin is a good quantum number in this system. In the
absence of orbital pair breaking \cite{note1} and spin-orbit
scattering, the Zeeman splitting of the localized Cooper pairs seems
to be the likely mechanism behind the suppression of the
Bose-insulator phase. At very high fields the behavior is again
metallic and independent of $B$, however with a zero-temperature
resistance, which is significantly different from $R_N$ at high $T$
and $B=0$. The resistance $R_\mathrm{sat}(T\rightarrow 0)=h/e^2$ in
this metallic phase turns out to be universal in the sense that it
is independent of the material and the degree of disorder in the
films.

A similar negative magnetoresistance with a saturation near $h/e^2$
in high magnetic fields has recently been reported for {\em
insulating} Be films by Butko and Adams~\cite{Butko}. Since the
saturation resistance appeared to be of purely quantum nature, these
authors have introduced the term ``quantum metallicity'' for the
peculiar metallic behavior at high magnetic fields. Their highly
disordered films have resistances up to $\sim 4 h/e^2$ in zero
magnetic field and do not reveal a zero resistance state down to
$T=40$~mK. Because of the strong similarities between their
insulating Be films and the superconducting InO$_x$ and TiN films
investigated here, we suggest to use the same term ``quantum
metallicity'' for the high magnetic field state, despite the evident
differences in zero magnetic field. It is well possible that the
insulating state in Ref.~\cite{Butko} is also formed by localized
Cooper pairs, since there are experiments demonstrating the
existence of the thickness- and the magnetic-field-tuned
superconductor-insulator transition also in thin
Be-films~\cite{Be3,Be1,Be2}.

These striking similarities of the high magnetic field behavior for
several different materials (Be, InO$_x$, and our TiN films),
showing the SIT, point towards a common microscopic mechanism
underlying the suppression of the Bose-insulator in the limit of
high magnetic field. In absence of any quantitative theory, we would
like to draw attention to the fact, that the observed strongly
negative magnetoresistance with a tendency to saturation can be
described by the remarkably simple empirical expression of
Eq.~(1)~\cite{remark}. This expression holds in a very wide range of
$B$ and $T$ and may prove useful for future theoretical
considerations required for a deeper understanding of the nature of
the Bose-insulator and the high-field quantum metallic state.

\begin{acknowledgments}
We thank V.~Gantmakher for providing the raw data from
Ref.~\cite{SITVFG} and continuous support, A.~Goldman for drawing
our attention to Ref.~\cite{Butko}, D.~Weiss and W.~Wegscheider for
access to their high magnetic field system, and M.~Feigel'man,
A.~Finkelstein, and V.~Vinokur for useful discussions.
This research has been supported by the
Program ``Quantum macrophysics'' of the Russian Academy of Sciences,
the Russian Foundation for Basic Research (Grant No. 06-02-16704),
and the Deutsche
Forschungsgemeinschaft within the GK 638.
\end{acknowledgments}


\begin{references}

\bibitem{Goldman_Review} For a review see e. g. A.~Goldman and
N.~Markovic, Physics Today {\bf 51}, No. 11, 39 (1998).

\bibitem{Fisher89} M.\,P.\,A.~Fisher and D.\,H.~Lee,
Phys. Rev. B {\bf 39}, 2756 (1989).

\bibitem{Fisher90_1} M.\,P.\,A.~Fisher, G.~Grinstein, S.\,M.~Girvin,
Phys. Rev. Lett. {\bf 64}, 587 (1990).

\bibitem{Fisher90_2} M.\,P.\,A.~Fisher, Phys. Rev. Lett. {\bf 65}, 923 (1990).

\bibitem{Sondhi} S.\,L.~Sondhi, S.\,M.~Girvin, J.\,P.~Carini, and D.~Shahar,
Rev. Mod. Phys. {\bf 69}, 315 (1997).

\bibitem{Dolgop} E.\,L.~Shangina and V.\,T.~Dolgopolov,
Phys. Usp. {\bf 46}, 777 (2003).

\bibitem{Haviland} D.\,B.~Haviland, Y.~Liu, and A.\,M.~Goldman,
Phys. Rev. Lett. {\bf 62}, 2180 (1989).

\bibitem{Liu} Y.~Liu, D.\,B.~Haviland, B.~Nease, and A.\,M.~Goldman,
Phys. Rev. B {\bf 47}, 5931 (1993).

\bibitem{Hebard} A.\,F.~Hebard and M.\,A.~Paalanen,
Phys. Rev. Lett. {\bf 65}, 927 (1990).

\bibitem{Shahar} G.~Sambandamurthy, L. W.~Engel, A.~Johansson, and D.~Shahar,
Phys. Rev. Lett. {\bf 92}, 107005 (2004).

\bibitem{Destr} V.\,F.~Gantmakher, M.\,V.~Golubkov, V.\,T.~Dolgopolov,
G.\,E.~Tsydynzhapov, A.\,A.~Shashkin, Pis'ma ZhETF {\bf 68}, 337
(1998). [JETP Lett. {\bf 68}, 363 (1998)].

\bibitem{SITVFG} V.\,F.~Gantmakher, M.\,V.~Golubkov, V.\,T.~Dolgopolov,
A.\,A.~Shashkin, G.\,E.~Tsydynzhapov,  Pis'ma ZhETF {\bf 71}, 693
(2000). [JETP Lett. {\bf 71}, 473 (2000)].

\bibitem{KapitulnikSIT} M.\,A.~Steiner, G.~Boebinger, and A.~Kapitulnik,
Phys. Rev. Lett. {\bf 94}, 107008 (2005).

\bibitem{OurTiN04} T.\,I.~Baturina, D.\,R.~Islamov, J.~Bentner,
C.~Strunk, M.\,R.~Baklanov, and A.~Satta, JETP Lett. {\bf 79}, 337
(2004).

\bibitem{granular} see e.g. H.\,M.~Jaeger, D.\,B.~Haviland, B.\,G.~Orr,
and A.\,M.~Goldman, Phys. Rev. B {\bf 40}, 182 (1989);
A.~Frydman, O.~Naaman, and R.\,C.~Dynes, Phys. Rev. B {\bf 66}, 052509 (2002).

\bibitem{finkelstein}
A. M. Finkelstein, Physica B {\bf 197}, 636 (1994).

\bibitem{Be3} E. Bielejec, W. Wu,
Phys. Rev. Lett. {\bf 88}, 206802 (2002).

\bibitem{B_m} Typically one has $B_\mathrm{m}=8-10\,$T for InO$_x$, but
only 4.6~T for the sample in Ref.~\cite{SITVFG}.

\bibitem{feigelman05}
M. A. Skvortsov and M. V. Feigel'man, Phys. Rev. Lett. {\bf 95},
057002 (2005).

\bibitem{Larkin_pc} A. I. Larkin, private communication.

\bibitem{suzuki}
 T. Susuki, Y. Seguchi, and T. Tsuboi, J. Phys. Soc.
Jpn. {\bf 69}, 1462 (2000).

\bibitem{note1}
Preliminary measurements in parallel magnetic field result in a
qualitatively similar behavior as in the perpendicular field
orientation. This indicates that the orbital pairbreaking, which
usually determines the upper critical field is of less importance in
our films.

\bibitem{Butko} V.\,Yu.~Butko and P.\,W.~Adams,
Nature {\bf 409}, 161 (2001).

\bibitem{Be1} E.~Bielejec, J.~Ruan, and W.~Wu,
Phys. Rev. Lett. {\bf 87}, 036801 (2001).

\bibitem{Be2} E.~Bielejec, J.~Ruan, and W.~Wu,
Phys. Rev. B {\bf 63}, 100502(R) (2001).

\bibitem{remark} We have also tried to fit the experimentally observed
decay of the resistance per square with a power law
($R-R_\mathrm{sat} \propto B^{-\alpha}$). In this case it was not
possible to fit the data at different temperatures with the same
exponent and reasonable values of $R_\mathrm{sat}$.

\end{references}
\end{document}